# Enhanced Interfacial Dzyaloshinskii−Moriya Interaction in annealed Pt/Co/MgO structures

Anni Cao,[†, ‡] Runze Chen,[†] Xueying Zhang,[†, §] Xinran Wang,[†] Shiyang Lu,[//] Shishen Yan,[//] Bert Koopmans,[‡] and Weisheng Zhao[*, †, §]

[†] Fert Beijing Institute, BDBC, School of Microelectronics, Beihang University, Beijing 100191, China

[‡] Department of Applied Physics, Institute for Photonic Integration, Eindhoven University of Technology, PO Box 513, 5600 MB Eindhoven, The Netherlands

[§] Beihang-Goertek Joint Microelectronics Institute, Qingdao Research Institute, Beihang University, Qingdao 266104, China

[//] School of Physics, State Key laboratory of Crystal Materials, Shandong University, Jinan 250100, China

Ⓢ*Supporting Information*

**ABSTRACT** The interfacial Dzyaloshinskii−Moriya interaction (iDMI) is attracting great interests for spintronics. An iDMI constant larger than 3 mJ/m$^2$ is expected to minimize the size of skyrmions and to optimize the DW dynamics. In this study, we experimentally demonstrate an enhanced iDMI in Pt/Co/X/MgO ultra-thin film structures with perpendicular magnetization. The iDMI constants were measured using a field-driven creep regime domain expansion method. The enhancement of iDMI with an atomically thin insertion of Ta and Mg is comprehensively understood with the help of *ab-initio* calculations. Thermal annealing has been used to crystallize the MgO thin layer for improving tunneling magneto-resistance (TMR), but interestingly it also provides a further increase of the iDMI constant. An increase of the iDMI constant up to 3.3 mJ/m$^2$ is shown, which could be promising for the scaling down of skyrmion electronics.

**INTRODUCTION**

The Dzyaloshinskii−Moriya Interaction (DMI) is an antisymmetric exchange interaction that appears at inversion asymmetric structures and which leads to chiral spin texture. In most of the magnetic thin films, the interfacial Dzyaloshinskii−Moriya Interaction (iDMI) is a dominated contribution of DMI. The iDMI is one of the key ingredients for magnetic skyrmions[1–4], which are topologically protected spin structures, and chiral domain-walls (DW)[5–7]. It has been intensively studied in the past few years, and it was reported to influence the DW spin structures[8] and their current-driven dynamics[5,6,9,10]. Moreover, the DMI is responsible for establishing and controlling the sizes of magnetic skyrmions[11]. These small chiral spin textures are promising to be potential information carriers in future non-volatile spintronic applications, due to their unique properties including propagation driven by ultralow current densities[12–14] and re-writability by spin-polarized currents[15]. Although some theoretical and experimental efforts have been devoted to unveil the mechanism of DMI, it is still elusive particularly in non-epitaxial sputtered thin films. In systems of interest for spintronic applications, a strong DMI is urgently needed to overcome the exchange interaction and destabilize the uniform ferromagnetic state. Therefore, manipulating DMI efficiently is a crucial task for the development of advanced memory devices[16].

Our previous study[17] has proven that insertion of a "dusting" Mg layer in Pt/Co/MgO system can prevent the deterioration of the Co/MgO interface during the deposition, and can facilitate a better crystallization for both ferromagnetic and insulating layer. As the MgO thickness of the Pt/Co/Mg/MgO multilayers increases, the iDMI strength will increase at first and then saturate. In this paper, we propose Ta as an alternative inserted material and explore the role of thermal annealing to further enhance the DMI. We compare experimental results with first principle calculations to explain why Ta, as the inserted layer, gives rise to a slightly higher iDMI energy than Mg does. We also experimentally unveiled a relationship between the DMI and thermal annealing. The effective DMI fields of annealed samples were quantified by analyzing domain-wall motion in the presence of an in-plane field. Annealing processes were conducted on Pt/Co/Ta/MgO samples with 3 different MgO thicknesses. All of the samples exhibit an annealing-temperature-dependent DMI, which firstly increases and tends to decrease in the end. To the best of our knowledge, this is the first report of a DMI constant for Pt/Co/MgO multilayers of over 3 mJ/m².

**SAMPLE PREPARATION AND BASIC CHARACTERIZATION**

We use magnetron sputtering at room temperature to deposit multilayers with composition Ta(3 nm)/Pt(3 nm)/Co(1 nm)/X(0.2 nm)/MgO(t)/Pt(5 nm), as shown in Figure 1a. The inserted layer X is designed to be Ta or Mg, while the MgO thickness $t$ varies from 0 to 2.0 nm. The substrate consists of a 500-μm thick Si wafer with a 300-nm thermal oxide layer. The base pressure of our ultrahigh vacuum deposition system is around $3 \times 10^{-8}$ mbar. The bottom 3 nm Ta was able to create a (111) texture of the lower Pt, which provides the proper conditions for establishing perpendicular magnetic anisotropy (PMA) for the whole structure[18]. The inserted layer X was used for protecting Co from excessive oxidation. Moreover, we hope to strengthen DMI through this layer. Samples with different MgO thicknesses were also prepared to examine the variance of DMI. The top Pt performed as a protective layer preventing from the film oxidation.

A sectional view by spherical aberration corrected Transmission Electron Microscope (TEM) is shown in Figure 1b for the Pt/Co/Ta/MgO (1.2 nm) sample. Referring to the nominal thickness, the borders of each layer can be easily recognized, except for the Co/MgO interface. The clear Pt lattice proves the success of milling by focused ion beam (FIB) and the high quality of our multilayers. We use an alternating gradient field magnetometers (AGFM) to confirm the perpendicular magnetization and characterize the magnetic properties of the samples with the two different inserted layers at room temperature. The hysteresis loops for perpendicular applied field are depicted in Figure 1c. They show that the saturation magnetizations of samples with Ta inserted are slightly higher than the group of samples inserted with Mg. For the analysis of the DMI, also the in-plane loops are required. Results thereof can be found in Figure S1 of the SI.

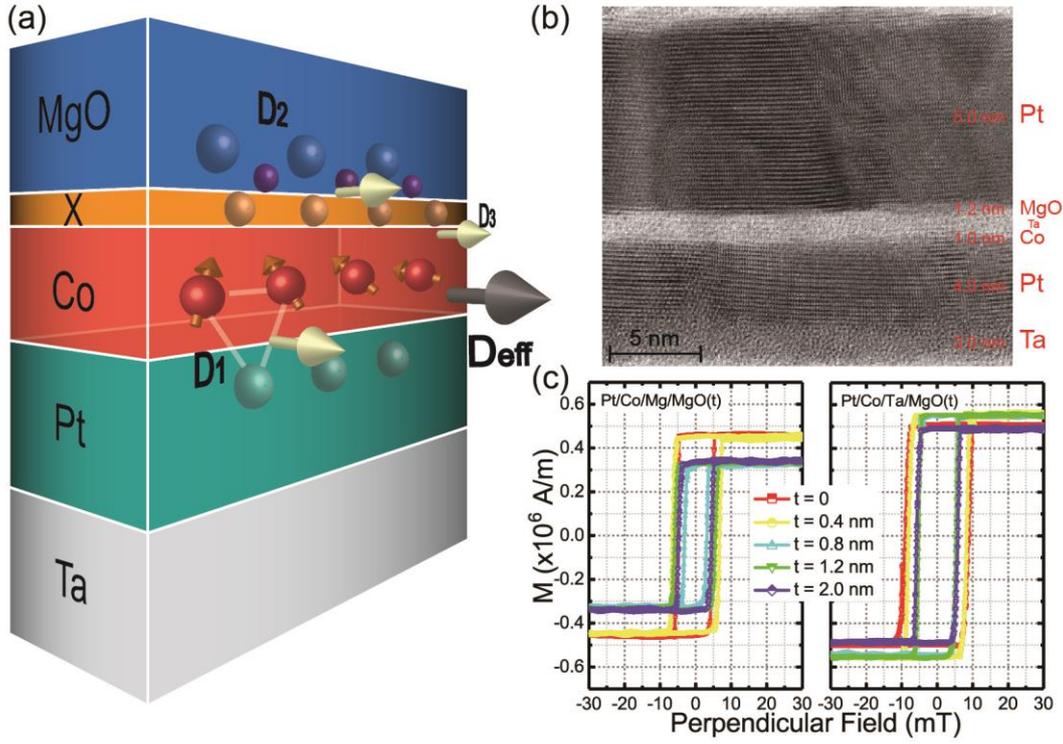

Figure 1. (a) Schematic of the Ta/Pt/Co/X/MgO stack structure. (b) Cross profile of the as-deposited Pt/Co/Ta/MgO sample with MgO thickness = 1.2 nm as measured by transmission electron microscopy. (c) Hysteresis loops with perpendicular applied field for Pt/Co/Mg/MgO(t) and Pt/Co/Ta/MgO(t) structures.

We quantified the strength of DMI in our samples, employing a Kerr microscope to observe asymmetric DW movement in the creep regime with an in-plane field $H_X$ and a perpendicular field $H_Z$. The dependence of DW velocities on the in-plane field is found to be roughly quadratic, where the minimum occurs at a non-zero value of $H_X$, which is defined as the effective DMI field $H_{DMI}$[19–21]. For typical examples see Figure S2 – S4 in the SI. From the DMI field $H_{DMI}$ one can extract the DMI constant

$$|D| = \mu_0 M_S |H_{DMI}| \sqrt{A/K_{eff}} \qquad (1),$$

using values of $M_S$ and $K_{eff}$ as obtained from the measured hysteresis loops, and $A$ from literature. For more details on the operation and analysis see our previous work [17]. Figure 2a exhibits experimental results of DMI for as-deposited samples inserted by Mg and Ta, with various MgO thickness. It can be deduced from Figure 2a that DMI almost remains unchanged with the increment of MgO thickness beyond a certain saturation level (1.2 – 2.0 nm here). Samples with Pt/Co/Ta/MgO structure and those with Pt/Co/Mg/MgO are found to have approximately the same saturation value of the DMI constant $|D|$, as shown in Figure 2a. First-principle calculations were adopted to judge the performance of Ta and Mg on DMI and to give a reasonable physical explanation.

**FIRST-PRINCIPLE CALCULATIONS**

To compare the DMI of the Pt/Co/MgO structure after inserting Mg and Ta, we used *ab initio* calculations on Co/X bilayers. One has to realize that the total DMI is a sum of contributions due to the Pt/Co and Co/X/MgO interface, as will be discussed later in more detail. From the calculations we extract the additional DMI from the Co/X/MgO interface. As a comparison, the DMI coefficient of the Co/MgO interface is also calculated. The DMI energy ($E_{DMI}$) can be depicted as

$$E_{DMI} = \sum_{<i,j>} d_{ij} \cdot (S_i \times S_j) \qquad (2),$$

where $S_i$ and $S_j$ are nearest neighboring normalized spins and $d_{ij}$ is the corresponding DMI vector. The total DMI strength $d^{tot}$ was calculated by identifying the difference between the clockwise energy $E_{CW}$ and anticlockwise energy $E_{ACW}$ (as defined in ref. 22) based on the density functional theory (DFT) for opposite chirality spin configurations with $E_{CW}$ and $E_{ACW}$ calculated from Eq. (2). A parameter $d^{tot}$ can be introduced according to $d^{tot} = (E_{CW} - E_{ACW})/12$[22]. The DMI strength can also be expressed by the micro magnetic energy per volume unit of the magnetic film with the corresponding coefficient $D^{tot}$. We can write $D^{tot}$ as $D^{tot} = \frac{3\sqrt{2} d^{tot}}{2 N_F r^2}$, in which $r$ is the distance between two nearest neighbor Co atoms and $N_F$ is the number of the magnetic atomic layers[22].

The VASP package[23,24] was employed using supercells with a monolayer (ML) of MgO on 3 ML of Co with the surface of MgO passivated by hydrogen, as well as a ML of Mg or Ta on 3 ML of Co (Figure 2b – 2d). In order to extract the DMI vector, calculations were performed in three steps. First, structural relaxations were performed until the forces become smaller than $0.001 \, eV/Å$ for determining the most stable interfacial geometries. Next, the Kohn-Sham equations were solved, with no spin orbit coupling (SOC), to find out the charge distribution of the system's ground state. Finally, SOC was included and the self-consistent total energy of the system was determined as a function of the orientation of the magnetic moments which were controlled by using the constrained method implemented in VASP[22].

It has been calculated in ref. [34] that the DMI energies at Pt/Co and Co/MgO interfaces are comparable, and we assume that the DMI energy at the Pt/Co interface is not affected by changing the inserted layer X. Considering the interfacial structure for multilayers grown by magnetron sputtering, the inserted monolayer is more likely not to be a closed layer but formed by a distribution of "islands" between Co and MgO. Figure 2b – 2d show the ideal interfacial atomic structure used for the DMI constant calculation, Co/MgO, Co/Mg and Co/Ta. Red, purple, blue and yellow spheres correspond to Co, O, Mg and Ta atoms respectively. In Figure 2(e), the total DMI coefficients $d^{tot}$ and the micro-magnetic DMI energy $D^{tot}$ for the three structures are compared. The values of Co/Mg (0.30 meV) and Co/Ta (-0.14 meV) are found to be much smaller than those of Co/MgO (1.86 meV). To some extent this indicates that bringing additional DMI is not the dominant function of the inserted layer. Rather, since the *ab-initio* calculations predict that DMI should reduce at atomic locations where a closed layer of X forms, this means that at places where the interface can be considered as Co/MgO-like (without X), *i.e.* at places where Co-O bonds dominate, the DMI should have increased. Therefore, we conjecture that insertion of Ta and Mg makes the pristine interface better, *i.e.*, it overcompensates the loss of DMI by the presence of Ta or Mg. Based on the opposite sign of $D^{tot}$ for Ta and Mg, this effect should be stronger for Ta than for Mg, since the calculated reduction of DMI for Ta is larger than for Mg. In this way, Ta is slightly better than Mg in enhancing the DMI energy of the Pt/Co/MgO system.

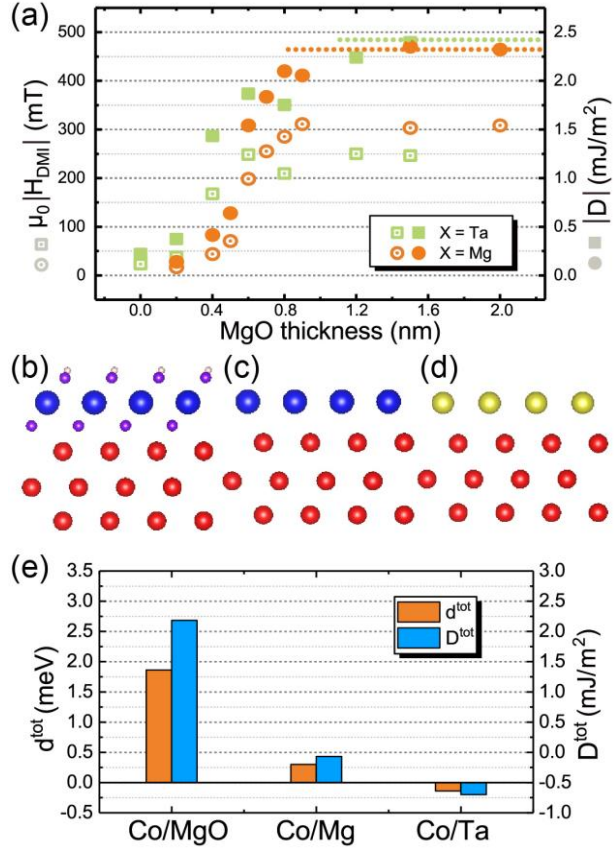

Figure 2. (a) Experimental trends of the effective DMI field $\mu_0|H_{DMI}|$ and DMI constant $|D|$ as a function of MgO thickness with different inserted layer X. Symbols with center dot represent $\mu_0|H_{DMI}|$ and solid symbols represent $|D|$. The ideal interfacial array of atoms used for the first principle calculation, (b) Co/MgO, (c) Co/Mg and (d) Co/Ta. (e) The total DMI strength $d^{tot}$ and the micro-magnetic DMI energy $D^{tot}$ of the three kinds of interfaces.

**ANNEALING EFFECTS ON MAGNETIC PROPERTIES**

Different samples were annealed for half an hour at temperatures ranging up to 380 °C, after which hysteresis loops and domain-wall motion was measured at room temperature. We controlled the rising rate and the duration of annealing temperatures to be the same, and applied a 50 mT perpendicular field while annealing. Hysteresis loops with perpendicular magnetic field for annealed samples with Ta inserted are shown in Figure 3a, for MgO thickness $t$ = 0.8, 1.2 and 1.5 nm. Corresponding in-plane field loops can be found in the SI as Figure S1. The sharp switching of the magnetization in the perpendicular loops are consistent with perpendicular anisotropy of all samples, although some details of the loops depend on thermal annealing. Figure 3b shows magnetic properties extracted from the hysteresis loops for samples with different annealing temperatures. We observe that the saturation magnetization $M_S$ of the sample with the thickest MgO shows hardly any dependence on annealing, whereas for the thinnest oxide sample there is a trend of an initial increase followed by a decrease. The effective anisotropy field $H_{k_{eff}}$ was obtained by extracting the field corresponding to 90% of the saturated magnetization in the hysteresis loops with in-plane magnetic field. As annealing temperature rise from 200 °C to 380 °C, $H_{k_{eff}}$ shrinks to 60% upon annealing. The effective magnetic anisotropy energy $K_{eff}$, calculated as $K_{eff} = \frac{1}{2}\mu_0 H_K M_S$[25,26], shows similar trends as $M_S$. With the increase of annealing temperature, the coercive field $H_C$ exhibits a 3 - 4 times growth compared with as-deposited samples, which is consistent with former studies[27]. Overall we find quite similar trends in the magnetic properties upon annealing Pt/Co/Ta/MgO samples with different MgO thickness.

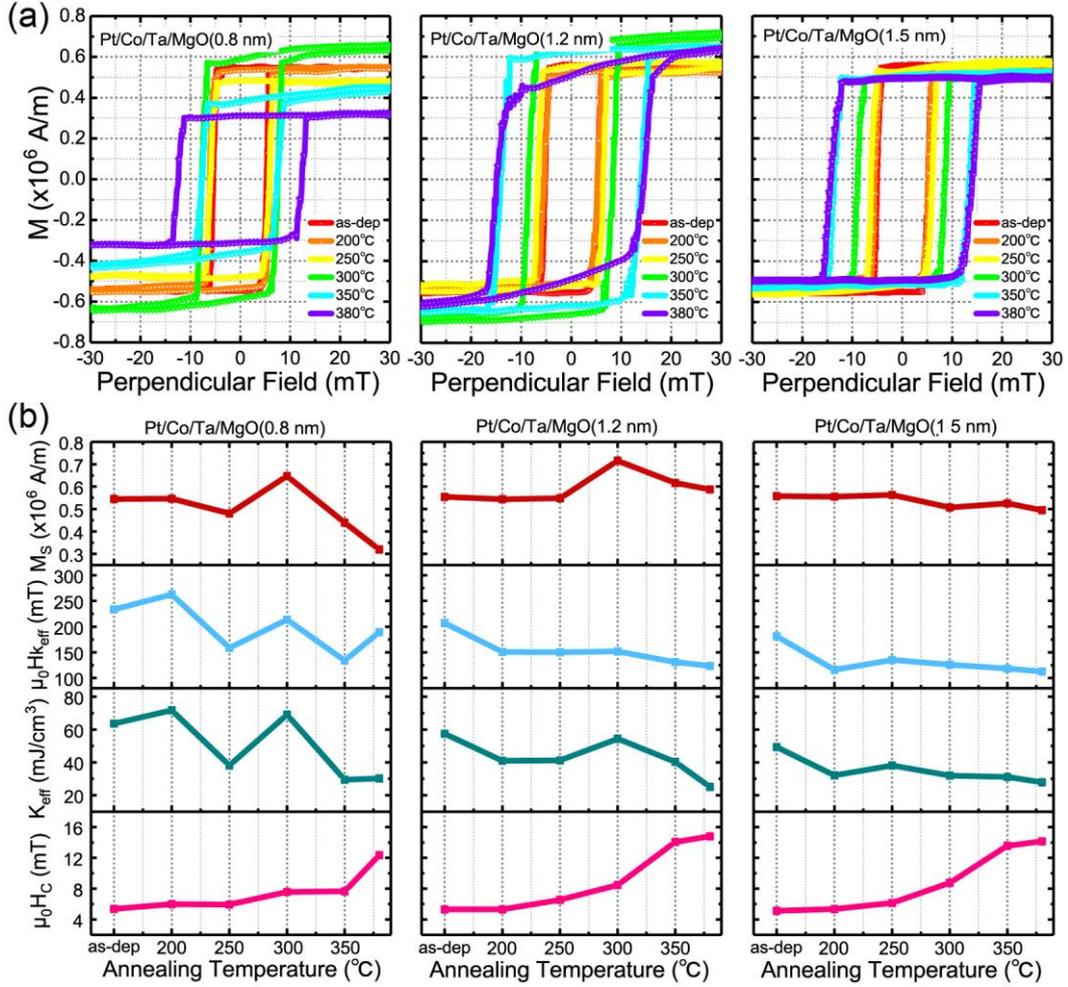

Figure 3. (a) Hysteresis loops applied with perpendicular field of annealed Pt/Co/Ta/MgO(t) structures while t = 0.8 nm, 1.2 nm and 1.5 nm. Different annealing temperatures of the samples with the same structure are distinguished by different color in each subfigure. (b) Magnetic properties obtained from the hysteresis loops for samples with different annealing temperatures. Subfigures in each line share common scales and y-axis' legends at left.

**ANNEALING EFFCTS ON DMI**

A typical result of, asymmetrical DW motions in the presence of an in-plane field for two directions of the out of plane field $\mu_0 H_z$ is shown in Figure 4a. The applied $\mu_0 H_z$ which varies from several milli-tesla to tens of milli-tesla for different samples, and the in-plane field $\mu_0 H_x$ was in the range of ±350 mT. A selection of our DW velocity measurements can be found in Figure S2 to Figure S4. We verified that domain-wall motion is in the creep regime for all perpendicular field applied, as shown in Figure S5 to Figure S7 in the supplementary material. Moreover, we carefully checked that the value of the extracted DMI field is independent of the magnitude of the applied perpendicular field, see Figure S8 in the SI. Thus, an unambiguous value of the DMI parameter is obtained for each sample at each annealing temperature. The absolute value of the DMI constant $|D|$ can be calculated by Eq. (1)[28]. By assuming the exchange stiffness constant $A = 15$ pJ/m[29,30], our $|D|$ can reach as high as 3.3 mJ/m$^2$. We are aware that the assessment of the exchange stiffness $A$ is not trivial, since the annealing process is quite possible to have an effect on it. Although its temperature dependence was directly ignored in some literatures[27,31], other work has suggested an increasing exchange stiffness increase of annealing temperature in similar thin films[32,33]. The later suggestion might mean that our estimate of DMI would be a conservative estimate, and its actual value would be higher. The effective DMI fields $\mu_0 |H_{DMI}|$ and DMI energy $|D|$ for three components Pt/Co/Ta/MgO(0.8 nm, 1.2 nm and 1.5 nm) assuming a constant $A$ are depicted in Figure 4b. We thus found that the strength of DMI manifests differences for different MgO thickness, but they all exhibit a trend of an initial increase followed by a decrease. Within the investigated range, the DMI values display an optimum at an annealing temperature of around 300 °C, independent of MgO thickness.

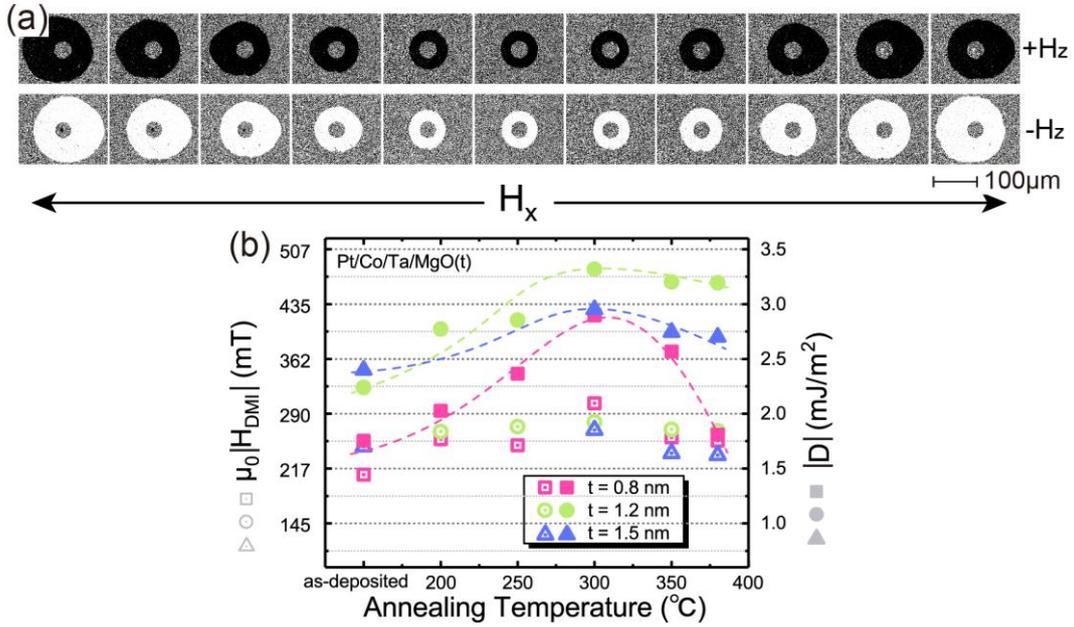

Figure 4. (a) DW expansion of the same sample driven by an out-of-plane magnetic field $\mu_0|H_z| = 10.62$ mT and a varying in-plane field $\mu_0 H_x$. (b) Trends of the effective DMI field and DMI constant as a function of annealing temperature. Square symbols, circular symbols and triangle symbols stand for the MgO thickness $t$ = 0.8 nm, 1.2 nm and 1.5 nm separately. Symbols with center dot stand for $\mu_0|H_{DMI}|$ while those solid ones stand for $|D|$.

**DISCUSSION**

In the Pt/Co/MgO system, the large interfacial DMI $iDMI_{Pt/Co/MgO}$ does not only come from the strong SOC between the Pt and Co, but also has a significant contribution from the Co/MgO interface, following the expression $iDMI_{Pt/Co/MgO} = iDMI_{Pt/Co} + iDMI_{Co/MgO}$. The DFT calculations have proven that $iDMI_{Pt/Co}$ and $iDMI_{Co/MgO}$ have the same sign[34]. It has been accepted that interfacial oxidation is related to large charge transfer and to the large interfacial electric field that compensates the small spin–orbital coupling of the atoms at the interface, which directly increase the DMI[35,36]. The inserted X layer efficiently protects the Co layer from degradation, and a proper material could strengthen the asymmetry of the whole structure, and consequently enhance the DMI.

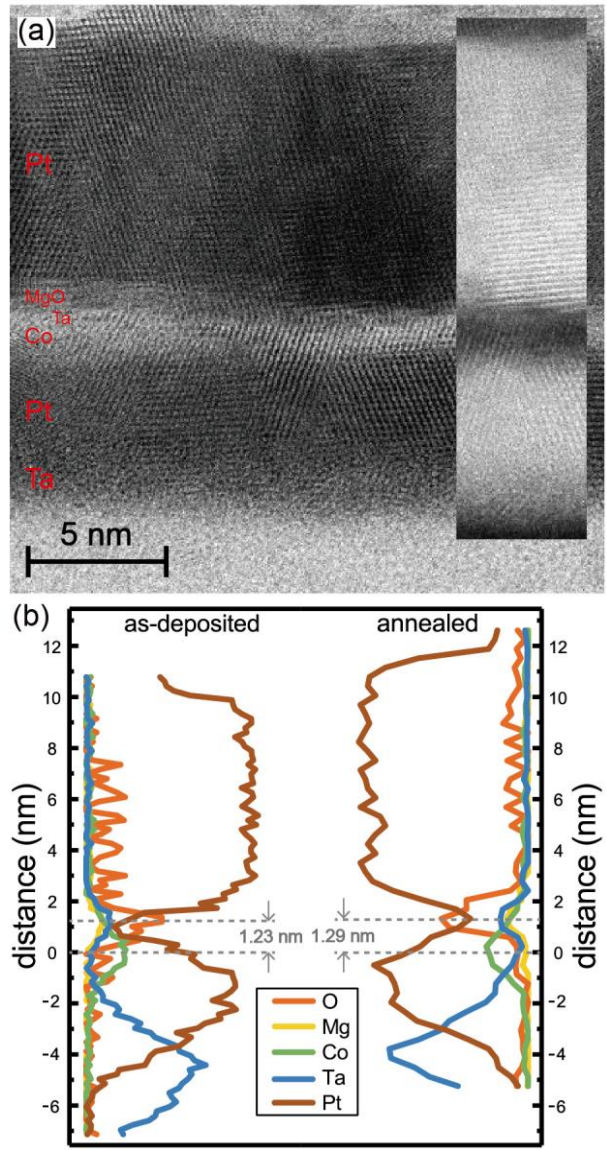

Figure 5. (a) Cross profile of the Pt/Co/Ta/MgO (1.2 nm) sample after annealing at 300 °C as measured by transmission electron microscopy. The inserted subfigure is the result with an inverting imaging field. (b) X-ray energy dispersive spectroscopy curves of the as-deposited sample and the 300 °C annealed sample.

It was reported that the annealing process would homogenize the oxide layer[37,38], and improved the Co/MgO interface, though there is not a layer X between Co and MgO in former studies. To confirm this, another TEM image is given in Figure 5a. Compared with Figure 1b, the degrees of crystallinity for Co and MgO layers are appreciably improved. We also exhibit a comparison of X-ray energy dispersive spectroscopy (EDS) curves for the Pt/Co/Ta/MgO (1.2 nm) sample before and after 300 °C thermal annealing in Figure 5b, where the curves are shifted such that the Co peak positions defined the zero position of the scan. Thus we see that the distributions of each element are not changed too much by annealing. We find that the O atoms' peak position shows a small, but finite 5% shift for the annealed sample, which would be consistent with the slight growth of $M_S$ for the 300 °C annealed sample (seen as Figure 3e). Secondly, an improved ordering of the atoms at the Pt/Co interface, which is brought about by annealing, might be another reason for the initial enhancement of the DMI, since the DMI is sensitive to the atomic arrangements at the interface[10,39]. Following the increasing trend, a higher temperature will prompt the formation of a CoPt alloy at the Pt/Co interface and reduces the number of Co-O bonds[38]. Furthermore, it was reported[22] that annealing at higher temperatures leads to interfacial diffusion, being detrimental for the DMI. Therefore, a decreasing trend of DMI appears when the temperature goes above 300 °C. The similar trend was also found in Ta/CoFeB/MgO tri-layers[40]. Above all, the non-monotonic trend of DMI can be explained rationally.

The Pt/Co/MgO structure we studied here is very similar to the configuration of a tunnel barrier layer/free layer/capping layer of the most popular MTJ structure. Thermal annealing is a necessary way to produce a crystallized MgO tunnel barrier, thus improving the tunneling magnetoresistance (TMR) effect in magnetic multilayers[41]. Therefore, our study will be very relevant for applications that make use of electrical detection of magnetic skyrmions through TMR in MTJ devices. Further improvement of the lattice on asymmetric interface by thermal annealing is an essential way to fine-tune the DMI in Pt/Co/MgO samples which is valuable for the induction of chiral magnetic order.

**CONCLUSION**

In summary, by a combined experimental and theoretical study prove that insertion of both X = Ta and Mg in Pt/Co/X/MgO improves DMI significantly, while the effect on the interface quality may be slightly better for Ta than for Mg. Furthermore, we investigated the effect of thermal annealing on the DMI. A significantly enhanced iDMI is found in our annealed Pt/Co/Ta/MgO system. The optimal condition for the Pt/Co/Ta/MgO structure is found to be annealing around 300 ℃, for 0.5 hour, enhancing DMI to the largest extent. The influence of annealing is attributed to both Pt/Co and Co/MgO interface transformation. Our study will significantly contribute to research that relies on strong DMI in thin film systems, and to stabilize magnetic skyrmions at room-temperature.

**ASSOCIATED CONTENT**

Additional experiment results of DW motion can be seen as the **Supporting Information** (SI).

**AUTHOR INFORMATION**
**Correspongding Author**
*E-mail: weisheng.zhao@buaa.edu.cn (W.-S.Z.).

**Acknowledgment.** The authors would like to thank the supports by the projects from National Natural Science Foundation of China (No.61627813 and 61571023), the National Key Technology Program of China 2017ZX01032101, the Program of Introducing Talents of Discipline to Universities in China (No. B16001), the VR innovation platform from Qingdao Science and Technology Commission and the China Scholarship Council.